\documentclass[pdftex,twocolumn,epjc3]{svjour3}          

\RequirePackage[T1]{fontenc}

\smartqed  

\RequirePackage{graphicx}
\usepackage{latexsym}
\usepackage{amssymb, amsmath}
\usepackage{subfig}
\RequirePackage{mathptmx}      
\RequirePackage{flushend}
\RequirePackage[numbers,sort&compress]{natbib}
\RequirePackage[colorlinks,citecolor=blue,urlcolor=blue,linkcolor=blue]{hyperref}

\newcommand{\fr}{\mathtt{R}}
\newcommand{\fl}{\mathtt{L_m}}
\newcommand{\frl}{f(\mathtt{R,L_m})}

\journalname{Eur. Phys. J. C}

\begin{document}

\title{Massive white dwarfs in $f(\mathtt{R,L_m})$ gravity}


\author{R. V. Lobato\thanksref{e1, addr1, addr2}
  \and
  G. A. Carvalho\thanksref{e2, addr3}
  \and
  N. G. Kelkar\thanksref{e3, addr1}
  \and
  M. Nowakowski\thanksref{e4, addr1}
}

\thankstext{e1}{e-mail: r.vieira@uniandes.edu.co}
\thankstext{e2}{e-mail: araujogc@ita.br}
\thankstext{e3}{e-mail: nkelkar@uniandes.edu.co}
\thankstext[$\star$]{e4}{e-mail: mnowakos@uniandes.edu.co}

\institute{Departamento de Fisica, Universidad de los Andes, Bogot\'a, Colombia\label{addr1}
  \and Department of Physics and Astronomy, Texas A\&M
  University-Commerce, Commerce, TX 75429, USA\label{addr2}
  \and Instituto de Pesquisa e Desenvolvimento (IP\&D), Universidade do Vale do Para\'iba, 12244-000, S\~ao Jos\'e dos Campos, SP, Brazil\label{addr3}
}

\date{Received: date / Accepted: date}

\maketitle
\begin{abstract}
In this work, we investigate the equilibrium configurations of massive white dwarfs (MWD) in the
  context of modified gravity, namely $\frl$ gravity, where $\fr$ stands for the Ricci scalar
  and $\fl$ is the Lagrangian matter density. We focused on the specific case
  $\frl = \fr/2 + \fl + \sigma \fr\fl$, i.e., we have considered a non-minimal coupling between the gravity field
  and the matter field, with $\sigma$ being the coupling constant. For the first time, the theory is
  applied to white dwarfs, in particular to study massive white dwarfs, which is a topic of great interest in the last years. The equilibrium configurations predict maximum masses which are above the
  Chandrasekhar mass limit. The most important effect of the theory is to increase significantly the mass for stars
  with radius < 2000 km. We found that the theory can accommodate the super-Chandrasekhar white
  dwarfs for different star compositions. Apart from this, the theory recovers the General Relativity results for
  stars with radii larger than 3000 km, independent of the value of $\sigma$.
\end{abstract}

\section{Introduction}
\label{sec:int}
Recently, the possibility of some white dwarfs (WD) surpassing the Chandrasekhar mass limit was
raised~\cite{andrewhowell/2006,hicken/2007, yamanaka/2009, scalzo/2010, taubenberger/2011,
  silverman/2011}. The possible progenitor of such massive stars ($M>2 M_\odot$) are being termed
super-Chandrasekhar white dwarfs. In this context, plenty of works have been done to model these very
massive white dwarfs (MWD). The properties of MWDs have been
studied in different contexts: rotation
\cite{Boshkayev2011Dec, Boshkayev2012Dec, Franzon2015Oct, becerra/2019}; high electric and
magnetic fields \cite{coelho/2014, coelho/2014a, Liu2014May, lobato/2015, lobato/2015a, subramanian/2015, das/2015b, lobato/2016a, Bera2016Mar,
  malheiro/2017, lobato/2017, lobato/2017a, otoniel/2017, Chatterjee2017Jul, Carvalho2018May, Otoniel2019Jul};
temperature \cite{caceres/2017, nunes/2021} and modified
gravity~\cite{das/2015, jain/2016, carvalho/2017, banerjee/2017, vasconcellos/2018, saltas/2018, liu/2019,
  eslampanah/2019, chowdhury/2019, sun/2020, allahyari/2020, Rocha2020May}. It
has been shown that general relativistic effects are essential for massive white
dwarfs and cannot be disregarded \cite{cohen/1969, Carvalho2015Dec,
  Carvalho2016Apr, Mathew2017May, carvalho/2018b}. The main contribution of the
general relativistic effects is in the radius of the stars, which,
as pointed out in Ref. \cite{Carvalho2015Dec} can be up to 37\% for some cases. General relativistic effects are also important for the stellar structure, since they can lead to global instabilities, limiting the maximum mass.
Modified gravity is also {\it per se} a
topic of great current interest due to the unknown entity (dark energy)
which accelerates~\cite{riess/1998} the
Universe. Therefore, through the modified theory of gravity, many strategies were developed to explain the
rapid expansion~\cite{nojiri/2017}.

White dwarfs have been useful to constrain parameters in modified theories~\cite{jain/2016,olmo/2020,Banerjee2017Oct}. That is
possible due to the well-known equation of state (EoS)
that describes the white dwarfs, as well as the huge amount of
observational data for these stars. This makes white dwarfs also a good “laboratory” to test the strong gravity regime.

In the present work, we are continuing to investigate compact objects in $\frl$ gravity, which we
have applied in previous works to neutron stars~\cite{carvalho/2020a, lobato/2021}. In those works, we showed that this theory can account for the enhancement of the maximum mass in neutron stars, in
better agreement with the observational data from GW170817 and {\it NICER} as compared to General Relativity~\cite{lobato/2021}. This theory is a generalization of the so-called $f(\fr)$ theories and was proposed by Harko and
Lobo~\cite{harko/2010}. The $\frl$ theory considers an explicit coupling between matter and geometry, described by a general function that depends on the Ricci scalar $\fr$ and the matter Lagrangian
density, $\fl$ i.e., there is a term such as $\sigma\fr\fl$ in the Lagrangian, where $\sigma$ is a coupling constant. Models with non-minimal
curvature-matter coupling have been extensive objects of investigation \cite{bertolami/2008}. As stated in Ref. \cite{harko/2010}, as a result of the coupling, the motion of particles is non-geodesic, there is an extra force orthogonal to the four-velocity and so on. The study and viability of the theory already started in different
contexts~\cite{harko/2010a, harko/2012, barrientos/2018} and the results are promising.

In the next section, we shall discuss the possible progenitor candidates for
supermassive white dwarfs, showing the present and future observations that lead
to them. In section \ref{frl} we shall briefly present the resulting hydrostatic equilibrium equation
for the $\frl$ theory of gravity. In Sec. \ref{numerical}, we will present the numerical procedure,
boundary conditions, the equation of state and threshold due to the
composition used in this work. Results are presented in Sec. \ref{results}, followed by the discussion and conclusions in the last section, \ref{conclu}.

\section{Super-Chandrasekhar candidates}
Though the origin of the supernovas of type Ia (SN Ia) is understood as the thermonuclear
detonation of a white dwarf (WD) which is triggered either by the merger of two WDs
in a binary or a single WD accreting mass from a companion, the experimental evidence
for the progenitors of SN Ia is scarce. Observation of progenitors near and above the
Chandrasekhar limit is challenging. In particular, some observed SN Ia, such as SN 2009dc, have spectral observations similar to normal SN Ia, however they are overluminous and ejecta velocities are slower compared to normal ones \cite{andrewhowell/2006,hicken/2007,yamanaka/2009,scalzo/2010}. The ejecta masses are estimated to be highly super-Chandrasekhar (2.2-2.8 $M_\odot$) \cite{taubenberger/2011}; while double degenerate merger, off-center explosion and differential rotation models still struggle to explain these large values \cite{Chen2009Aug,Tanaka2010Apr,Hachinger2012Dec}.

Besides the highly super-Chandrasekhar SN Ia progenitors, spectra of the pulsating subdwarf-B (sdB) star KPD 1930+2752 confirmed that this star is a binary \cite{maxted/2000}. The amplitude of the orbital motion (349.3 $\pm 2.7 {\rm km s^{-1}}$) combined with the canonical mass for sdB stars (0.5 M$_{\odot}$) implied a total mass for the binary of 1.47 $\pm$ 0.01 M$_{\odot}$, thus making it the first possible candidate for a super-Chandrasekhar progenitor. Another discovery of the shortest period binary comprising a hot subdwarf star
(CD-30 11223, GALEX J1411-3053) and a massive unseen companion was reported in
\cite{vennes/2012}. The measured parameters of the sdB CD-30 11223 were found to favor
a canonical mass close to 0.48 M$_{\odot}$ that would correspond
to a minimum mass of 0.77 M$_{\odot}$ for the companion.
Constraining the radius of the primary using a measurement of its rotation
velocity and fitting the amplitude of the ellipsoidal variations, the authors set an
upper limit of 0.87 M$_{\odot}$ for the secondary mass at an
orbital inclination of 68$^o$ and hence obtained a total mass of 1.35
M$_{\odot}$ for the system. However, the authors mention that
systematic effects in the measurement of the rotation velocity and the possibility
that the primary mass may exceed 0.48 M$_{\odot}$ imply that the total mass of the
system may exceed the Chandrasekhar mass limit.

More recently, HD265435, a binary system with an orbital period of less than a
hundred minutes, consisting of a white dwarf and a hot subdwarf was discovered
\cite{pelisoli/2021}. This system was observed by the Transiting Exoplanet Survey
Satellite (TESS). Combining the spectra obtained at the Palomar 200-inch telescope
with the radius estimate from fitting the spectral energy distribution (SED),
the visible star was characterized to be a hot subdwarf of spectral type OB.
The orbital inclination of the system allowed the authors to estimate the mass of
the unseen companion, which is probably a white dwarf with a carbon-oxygen core.
The total mass of the system was estimated to be 1.65 $\pm$ 0.25 solar masses, thus
exceeding the maximum allowed value of the mass of a stable white dwarf to make it
a super-Chandrasekhar candidate progenitor.

\section{Massive white dwarfs on the grounds of modified
  gravity}\label{mods}
As we mentioned above, the Chandrasekhar limit may not be unique. Some early studies considered a high magnetic field to violate it. However,
it was soon shown that one can have a maximum limit for extreme
  magnetic fields in these highly magnetized WDs, i.e., huge magnetic
  fields bring instabilities to the hydrostatic equilibrium equations.
For details and
  discussions see Ref. \cite{coelho/2014a} and
  references therein. Soon after the stability of these magnetic white
  dwarfs was explored in solid grounds, and a maximum magnetic field
  was established, depending on the core composition of the star
  \cite{chamel/2013}. Considering all sorts of instabilities, the
  maximum limit was established around $2.0\ M_{\odot}$
  \cite{subramanian/2015, chatterjee/2017a, otoniel/2017, Otoniel2019Jul}.
Above this limit the star
  changes its geometry too much, becoming a torus due to anisotropic
  pressures. Simulations of mergers of WDs leading to magnetic
  super-Chandrasekhar WDs have been perfor\-med \cite{becerra/2018} leading
  to a maximum mass of $1.45\ M_{\odot}$. For the case of isolated sources, it
  would be $1.46\ M_{\odot}$ \cite{becerra/2019}. This is in accordance
  with new observations such as the one in Ref. \cite{caiazzo/2021},
  where a WD has a mass around $1.32\ M_{\odot}$, radius around 2140
  km and a possible magnetic field of 600-900 megagauss. Although
  magnetic fields could enhance the mass, strongly magnetic white
  dwarfs are still absent among detached white dwarf binaries that are younger
  than one billion years. This is in contrast with semi-detached
  binaries, such as Cataclysmic Variables \cite{schreiber/2021}, i.e.,
  according to these data, super-Chandrasekhar due to magnetic field would
  likely be formed by Cataclysmic Variables as progenitors.

  As one can see, the highly expected breaking of spherical symmetry and the absence of strong magnetic fields in isolated white dwarfs, leads to an open
  window for other mechanisms that could also violate the Chandrasekhar
  limit. One such mechanism would be modified gravity \cite{das/2015, jain/2016, carvalho/2017, banerjee/2017, vasconcellos/2018, saltas/2018, liu/2019,
  eslampanah/2019, chowdhury/2019, sun/2020, allahyari/2020,
  Rocha2020May, wojnar/2021a}. In Ref. \cite{das/2015}, the well-known
$f(\fr)$ gravity which is one of the most studied modified theory of
gravity was used. The maximum masses found in this work were between
$1.772-2.701\ M_{\odot}$. This theory was also explored in the
Palatini formalism \cite{sarmah/2022} and for this formalism the WDs
could have mass beyond $2\ M_{\odot}$ for high values of the theory's
parameter. Such works used perturbative approaches to find the equilibrium configurations of WDs in $f(\fr)$ gravity, but recently Ref. \cite{Nobleson2022Jan} showed that perturbative results for neutron stars differ from non-perturbative ones, which means that those perturbative WD approaches can be misleading. The authors of Ref. \cite{banerjee/2017} also considered the $f(\fr)$ gravity as well as: fourth order gravity
theories (FOG), Eddington inspired Born-Infeld gravity (EiBI) and
Scalar-Tensor-Vector gravity (STVG). They used WD as a tool to
constraint these alternative theories. For STVG, there was no
significant enhancement of the mass for any value of the theory's
parameter. For the case of the EiBI, for different positive values of
the parameter of the
theory, one could have a significant enhancement of the mass, however
for some value that could lead to huge (unrealistic) mass-radius. For
the case of FOG, one can have the same behavior, i.e., a large
enhancement of the mass-radius. Finally, for the case of $f(\fr)$
gravity, where the model was characterized by two parameters, the
maximum mass of white dwarfs was near the Chandrasekhar limit. Unlike
FOG and EiBI, one could have the mass near
$3.0\ M_{\odot}$ for a pair of parameter combinations. Massive gravity
theory also could count for enhancement of the maximum mass of white
dwarfs. In Ref. \cite{eslampanah/2019}, the authors using the
Rham-Gabadadze-Tolley like massive gravity have found WDs with maximum
mass of 1.41 to 3.41 $M_{\odot}$ with radius of 871 to 1168 km,
respectively. As we can see the massive white dwarfs have been heavily
studied under modified gravity in an attempt to have white dwarfs
above the Chandrasekhar limit, nevertheless, WDs also have been used as
tools to test these modified gravity theories\cite{jain/2016, banerjee/2017, allahyari/2020}, which in general, are used
broadly in neutron star astrophysics. The aim is to have bounds in
the theories' coupling constant and its behavior in the Newtonian limit. The EoS of WDs is well-defined, as well as the
nuclear instabilities present in these systems and the mass-radius in
the Newtonian limit.

\section{Hydrostatic equilibrium equation in $\frl$ theory}
\label{frl}
The $\frl$ gravity action reads~\cite{harko/2010} as,

\begin{equation}\label{action}
S=\int d^{4}x\sqrt{-g}\, \frl,
\end{equation}
where $\frl$ is a general function of the Ricci scalar $\fr$ and of the
matter Lagrangian density $\fl$, $g$ is the metric determinant. When the function
takes the form $\frl = \fr/2 + \fl$, the principle of least action leads to the
Einstein's field equations $G_{\mu\nu}=T_{\mu\nu}$, where $G_{\mu\nu}$ is the Einstein tensor and $T_{\mu\nu}$ represents the energy-momentum tensor. We have been considering $c = 8\pi G =1$.

For the case where the function is $\frl = \fr/2 + \fl + \sigma \fr\fl$ as considered in references~\cite{garcia/2010,garcia/2011}; where $\sigma$ is the coupling constant, and $\fl= -p$; the variation of the action leads to the following field equations,
\begin{eqnarray}\label{fieldeqs}
  &&(1-2\sigma p)G_{\mu\nu}+\frac{1}{3} \fr g_{\mu\nu}-\frac{\sigma p}{3}
\fr g_{\mu\nu}\nonumber \\
   &-& (1+\sigma
  \fr)\left(T_{\mu\nu}-\frac{1}{3}Tg_{\mu\nu}\right) + 2\sigma \nabla_\mu\nabla_\nu p = 0.
\end{eqnarray}
which for the static spherically symmetric spacetime and taking the energy-momentum tensor for a
perfect fluid, leads to the Tolman-Oppenheimer-Volkov (T.O.V.) -
like equations, i.e., the hydrostatic
equilibrium equations,

\begin{subequations}\label{tov-like}
  \begin{eqnarray}
    \alpha'(p+\rho) + 2z = 0,
    \end{eqnarray}
    \begin{eqnarray}
      p' - z = 0,
    \end{eqnarray}
    \begin{eqnarray}
      &\Bigg[\bigg(2 r^{2} \rho e^{\beta} + \big(2  \fr r^{2} \rho e^{\beta} + 3 r^{2} z
             \alpha'+ 6  p r \beta' \nonumber \\
             &+ 2  \big(2  \fr p r^{2} + 3  p\big) e^{\beta} - 6
        p\big) \sigma - \left({\left(\fr - 3  p\right)} r^{2} + 3\right) e^{\beta} \nonumber \\
        &- 3  r \beta' +
        3\bigg) e^{-\beta}\Bigg]({3r^{2}})^{-1} = 0,
    \end{eqnarray}
    \begin{eqnarray}
     & \Bigg[\bigg(r^{2} \rho e^{\beta} + (\fr r^{2} \rho e^{\beta} + 3 \, r^{2} z \beta' + 6
       \, p r \alpha' - 6 \, r^{2} z' \nonumber \\ &- (\fr p r^{2} + 6 \, p) e^{\beta} + 6 \, p) \sigma + {\left(\fr r^{2} + 3\right)} e^{\beta} \nonumber \\
       &- 3 \, r
       \alpha' - 3\bigg) e^{-\beta}\Bigg]({3 \, r^{2}})^{-1} = 0.
\end{eqnarray}
\end{subequations}

Here, $\alpha$ and $\beta$ are the metric potentials depending on the
radial coordinate $r$ and primes denote their derivatives. Finally,
$p$ and $\rho$ are the pressure and energy density,
respectively, and $z$ is an auxiliary variable which is
the derivative of the pressure. For complete details, see
Ref.~\cite{lobato/2021}. It is worth pointing out that the trace
  of Eq. \eqref{fieldeqs} will provide $ \fr = -T$ for $\sigma=0$, and
  using this result one can recover the field equations of General
  Relativity for this particular case. Furthermore, one can note from
\eqref{tov-like}, that the four-divergence of the energy-momentum
tensor is conserved, which is a remarkable feature of the $\frl$
theory for stars with spherical symmetry. The four-divergence
  conservation of $T_{\mu\nu}$ is a consequence of our choice for the
  matter Lagrangian \cite{carvalho/2020a,lobato/2021}, $\fl=-p$, which
  is consistent with the on-shell Lagrangian for relativistic perfect
  fluids \cite{Brown1993Aug}. Finally we mention that for $\fl=0$,
  i.e., the vacuum case in which we also have $T_{\mu \nu}=0$ and
  $p=0$, the new Einstein equations reduce to $G_{\mu \nu}+ \fr  g_{\mu
  \nu}/3=0$. This implies $\fr = 0$ and hence $G_{\mu \nu}=0$.  We are
then back to vacuum Einstein equations and as a result the tests
of the theory under discussion are in stars, compact objects and cosmology.

\section{Numerical procedure, boundary conditions and equation of state}\label{numerical}

To solve the system of equations~\eqref{tov-like} numerically, we need to set an equation of state and the boundary conditions: the latter reads, $p(0) = p_c$ and $\rho(0) =
\rho_c$ at the center of the star $(r=0)$, where $p_c$ and $\rho_c$ are the central pressure and
central energy
density. The stellar surface is the point at the radial coordinate $r=R$, where
the pressure vanishes, $p(R)=0$. For the metric potentials, we use $\beta(0)=0$ and
$\alpha(0)=1$. For the auxiliary variable $z$, we use $z(0)=0$, once $p(0)$ is a global maximum point. The total mass is contained inside the radius $R$, as measured by the gravitational field
felt by a distant observer. As the boundary condition is at $r=R$ (the
Ricci scalar vanishes at the surface). The continuity of the metric,
i.e., the connection conditions with the exterior Schwarzschild solution, requires
that \cite{wojnar/2016, sbisa/2020, olmo/2020}

\begin{equation}\label{mass}
  M = m(R) = \int_0^R4\pi r^2\rho(r)dr.
  \end{equation}

\subsection*{Equation of state}
The simplest EoS which describes the fluid properties
of WDs follows the model used for the relativistic Fermi gas of
electrons \cite{chandrasekhar/1931, chandrasekhar/1935}, which is called
the Chandrasekhar EoS. There are other equations of state for WDs that insert
some corrections into the Chandrasekhar EoS, but essentially, they are all based on the Fermi gas of electrons. Because of this, there is little uncertainty in the EoS for WDs, which is not the case for neutron stars. Some studies, considering massive white dwarfs, generalized
the Chandrasekhar model to account for the mass threshold in the ultra-relativistic limit. Here, we point out
the work of Chamel and Fantina~\cite{chamel/2015}, where the threshold for density and pressure are found to be increased
due to electron-ion interactions.

\begin{table}[h]
  \centering
\begin{tabular}{cll}
  \hline
  $^A_Z X$	& $p\ ({\rm dyn\ cm^{-2}})$ & $\rho\ ({\rm g\
                                              cm^{-3}})$ \\
  \hline\hline
  $^4$He		& $3.59\times 10^{29}$ & $1.44\times 10^{11}$ \\
  $^{12}$C		& $6.99\times 10^{28}$ & $4.20\times 10^{10}$ \\
  $^{16}$O		& $2.73\times 10^{28}$ & $2.07\times 10^{10}$ \\
  $^{20}$Ne		& $6.21\times 10^{27}$ & $6.89\times 10^{9}$ \\
\hline
\end{tabular}
\caption{Pressure and density threshold for which matter becomes
  unstable for four elements. The
  pressure values are taken from
  Ref. \cite{chamel/2015}. Considering the threshold pressure within the
  Hamada-Salpeter EoS, we have the corresponding density threshold.}
\label{limits}
\end{table}

We use the Hamada-Salpeter (HS) EoS, which
accounts for corrections due to electrostatic energy, Thomas-Fermi
deviations, exchange energy and spin-spin interactions
\cite{hamada/1961, salpeter/1961}. However, only electrostatic corrections
are found to be non-negligi\-ble. This EoS changes the way that the
critical mass depends on the nuclear composition, i.e., now depends on
$A/Z$ and $Z$, while for Chandrasekhar it only has $A/Z$ dependence. The extended HS EoS slightly decreases the Chandrasekhar limit. Using the limit due to electron
capture instability, we constrain the massive
white dwarfs in $\frl$ gravity and the parameter of the theory. We use four elements: $^4$He,
$^{12}$C, $^{16}$O and $^{20}$Ne. The
maximum values for pressure are described in the table \ref{limits}, taken from
Ref. \cite{chamel/2015}. Using the HS EoS and the pressure thresholds of table \ref{limits}, we obtain the maximum allowed densities.

The Hamada-Salpeter equation of state can be written as \cite{salpeter/1961},
\begin{equation}
p = p_{\rm e} + p_{\rm l},
\end{equation}
\begin{equation}
\rho = \rho_{\rm e} + \rho_{\rm l} + \rho_{\rm i},
 \end{equation}
where,
\begin{subequations}
  \begin{eqnarray}
    p_{\rm e}
    = \frac{m_e^4c^5}{24\pi^2\hbar^3}f(x)
    \end{eqnarray}
    \begin{eqnarray}
    f(x) = x(2x^2 - 3)(x^2 + 1)^{1/2} + 3\sinh^{-1}x,
    \end{eqnarray}
    \begin{eqnarray}
p_{\rm l} &= -\frac{3}{10}\left(\frac{4\pi}{3}\right)^{1/3}Z^{2/3}e^2\frac{m_e^4c^4}{(3\pi^2)^{4/3}\hbar^4}x^4,
\end{eqnarray}
\end{subequations}
and
\begin{subequations}
  \begin{eqnarray}
    \rho_{\rm e}
    = \frac{m_e^4c^5}{8\pi^2\hbar^3}g(x),
    \end{eqnarray}
    \begin{eqnarray}
    g(x) = x(2x^2 + x)(x^2 + 1)^{1/2} - \sinh^{-1}x,
    \end{eqnarray}
    \begin{eqnarray}
\rho_{\rm l} &= -\frac{9}{10}\left(\frac{4\pi}{3}\right)^{1/3}Z^{2/3}e^2\frac{m_e^4c^4}{(3\pi^2)^{4/3}\hbar^4}x^4,
      \end{eqnarray}
      \begin{eqnarray}
\rho_{\rm i} &= m_Nc^2\frac{A}{Z}\frac{m_e^3c^3}{3\pi^2\hbar^3}x^3.
      \end{eqnarray}
\end{subequations}
 The subscripts e, l and i denote the degenerate electrons (Chandrasekhar EoS) term, the Coulomb interactions in the lattice and the rest-mass energy of the ions terms, respectively. $x$ is the relativity parameter defined in
terms of the Fermi momentum ${\rm k_f}$ as $x \equiv {\rm k_f}/mc$.

\section{Results}\label{results}
\begin{figure}[h]
  \includegraphics[scale=0.57]{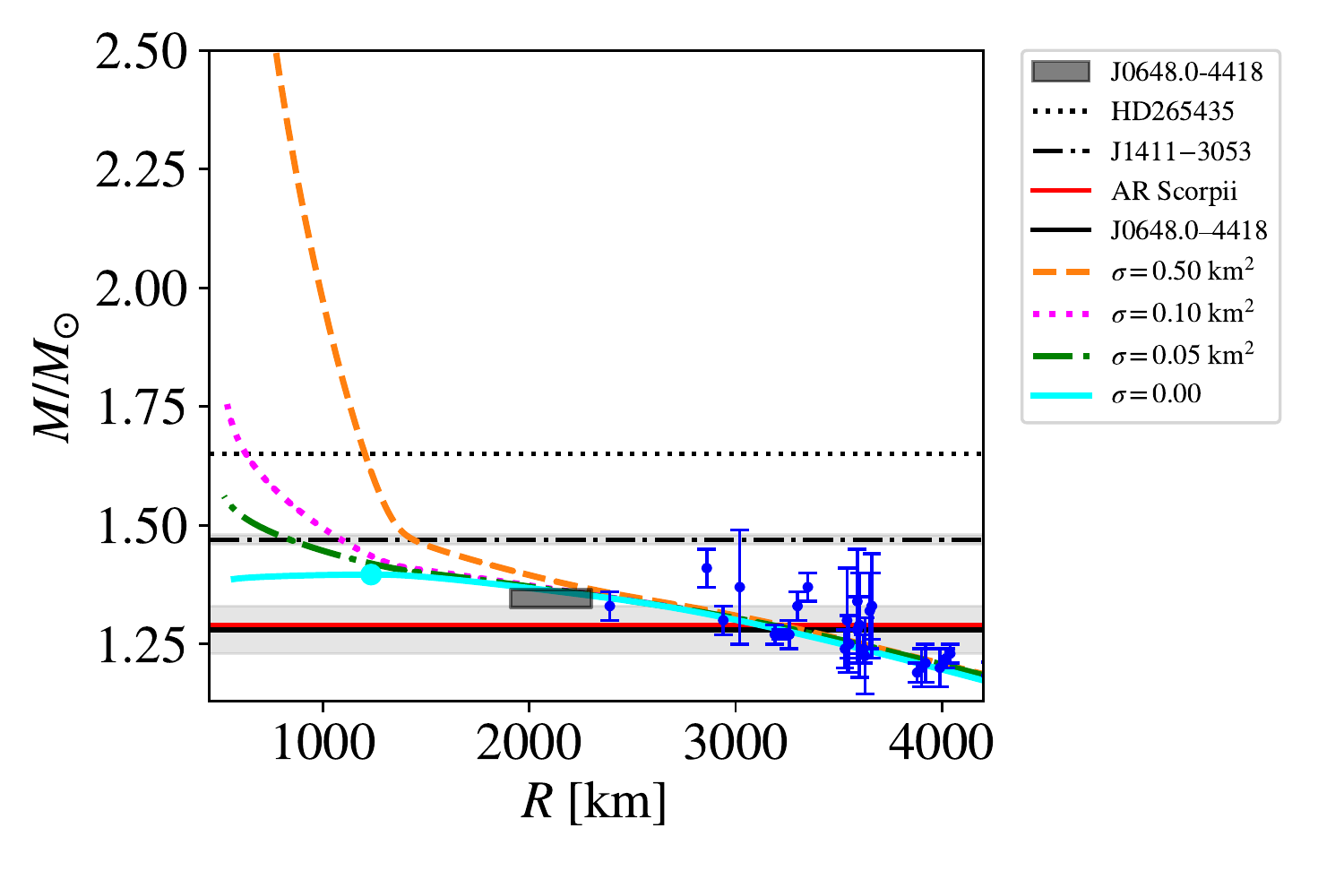}
\caption{Mass radius relationship for white dwarfs with different $\sigma$ parameters and
  considering a star made of Helium. Four values of $\sigma$ were considered. For $\sigma=0.0$,
  the theory represents General Relativity. The black-shaded region indicates the observed mass-radius
  of ZTFJ1901+1458, this region
  and mass-radius are taken from Ref.~\cite{caiazzo/2021}.  The
    black line and shaded region are $1.28\pm0.05\ M_{\odot}$
    corresponding to RX J0648.0-4418 \cite{mereghetti/2011}. The red
    line represents the upper limit of the compact object in AR
    Scorpii \cite{marsh/2016}. The black dot-dashed line with shaded
    region is the binary system with total mass of $1.47\pm0.01\
    M_{\odot}$ from Ref. \cite{maxted/2000} and the black dotted line is
  the binary system with a total mass of $1.65\pm0.25\ M_{\odot}$
  from Ref. \cite{pelisoli/2021}. The blue circles with error bars represent the
  observational data of a sample of massive WDs taken from
  Refs.~\cite{vennes/1997, nalezyty/2004}.}
\label{fig:he}.
\end{figure}

In figure~\ref{fig:he}, we present the mass-radius relationship for white dwarfs. We generated the
mass-radius within the $\frl$ gravity framework, considering four values for the coupling
constant $\sigma$. For $\sigma=0$, the theory recovers the General Relativity outcomes. For the other
cases, $\sigma$ is assumed to have positive values, going from $0.05$ to $0.5\
{\rm km^2}$. The constant presents
  different values from previous works, where it was larger for
  neutron stars \cite{carvalho/2020a, lobato/2021}, and smaller for weak-field limit
  \cite{harko/2010a, harko/2012}. As we pointed in our previous works, $\sigma$ has a dependence on the energy-matter density, i.e., depending on the
  astrophysical system, the parameter will have a different value. We expect that for black holes,
  the absolute values of the parameters will be the largest ones, while for the weak regime the smallest ones.
In figure~\ref{fig:he}, we have
considered the threshold for $^4$He. We have the maximum
pressure and central density: $p_c = 3.59\times
10^{29}\ {\rm dyn\ cm^{-2}}$ and $\rho_c = 1.44\times 10^{11}\
{\rm g\ cm^{-3}}$. As one
can see, the theory increases the mass in the diagram. The effects start to be more accentuated for stars with
radii $R< 2000$ km, i.e., for more massive stars the modified gravity effects
start to be non-negligible and the curves deviate from the GR
  regime. Considering the effects from the theory, it is possible to
surpass the Chandrasekhar limit of $1.4M_{\odot}$. For the value of
$\sigma=0.5\ {\rm km^2}$, the contribution of the theory becomes so relevant, that it is possible to have white dwarfs with more than two solar masses, if one
considers the $^4$He density threshold. These stars made of Helium and
considering $\frl$ gravity with $\sigma=0.5\ {\rm km^2}$, would be within the super-Chandrasekhar limit (2.1 -
2.8$M_{\odot}$), giving a possible explanation for the superluminous supernovae SNIa, since they will
be near instability. In the figure we have a black-shaded square region
which corresponds to the mass-radius observations of the white dwarf ZTFJ1901+1458 \cite{caiazzo/2021}. We also highlight massive WD observations in Fig. \ref{fig:he}; the blue dots with error bars, which are between
2500 and 4000 km \cite{vennes/1997, nalezyty/2004}. For the region $R>2500$ km, one can see that the contribution from the theory is almost negligible, even for the highest values of the coupling parameter $\sigma$. This behavior is entirely different from a previous
work, where we applied the $f(\fr,T)$ theory of gravity to white dwarfs,
e.g., see figure 2 of our work~\cite{carvalho/2017}. In this
  figure, we also have two systems: the white dwarf in the system HD
  49799 the RX J0648.0-4418 \cite{mereghetti/2011} with a mass of
  $1.28\pm0.05\ M_{\odot}$ (black line and shaded region) and the compact object in the AR Scorpii
  system \cite{marsh/2016}, with an upper limit of $1.29\ M_{\odot}$
  (red line). The former is a new system that has attracted
  attention in the last years with its being a possible white dwarf pulsar. We
also include two binary systems that could be possible super-Chandrasekhar progenitors, i.e., two possible candidates
for super-Chandrasekhar WDs in the future. The first system is the
J1411-3053 in black dot-dashed line with shaded region. This system has a total mass of
$1.47\pm0.01\ M_{\odot}$; the second one in black dotted-line is the system HD265435 with a
total mass of $1.65\pm0.25\ M_{\odot}$. In this system we have not
considered the error bar in the figure. For the curve $\sigma = 0$ we have a dot indicating the GR instability, i.e., one has $dM/dR>0$,
defining a maximum mass before the nuclear instabilities. When one
has the effects of the theory, this behavior changes. We have more
massive stars with a decrement of the central density, as in the case
of the $f(\fr,T)$ gravity \cite{carvalho/2017}. This is the same when
magnetic white dwarfs are considered. Due to this feature, the limiting
factor always will be the nuclear instabilities and that should be
treated carefully when one wants to constrain extended gravity
theories' parameters and find the WD's maximum mass in such a theory.

\begin{figure}[h]
\includegraphics[scale=0.552]{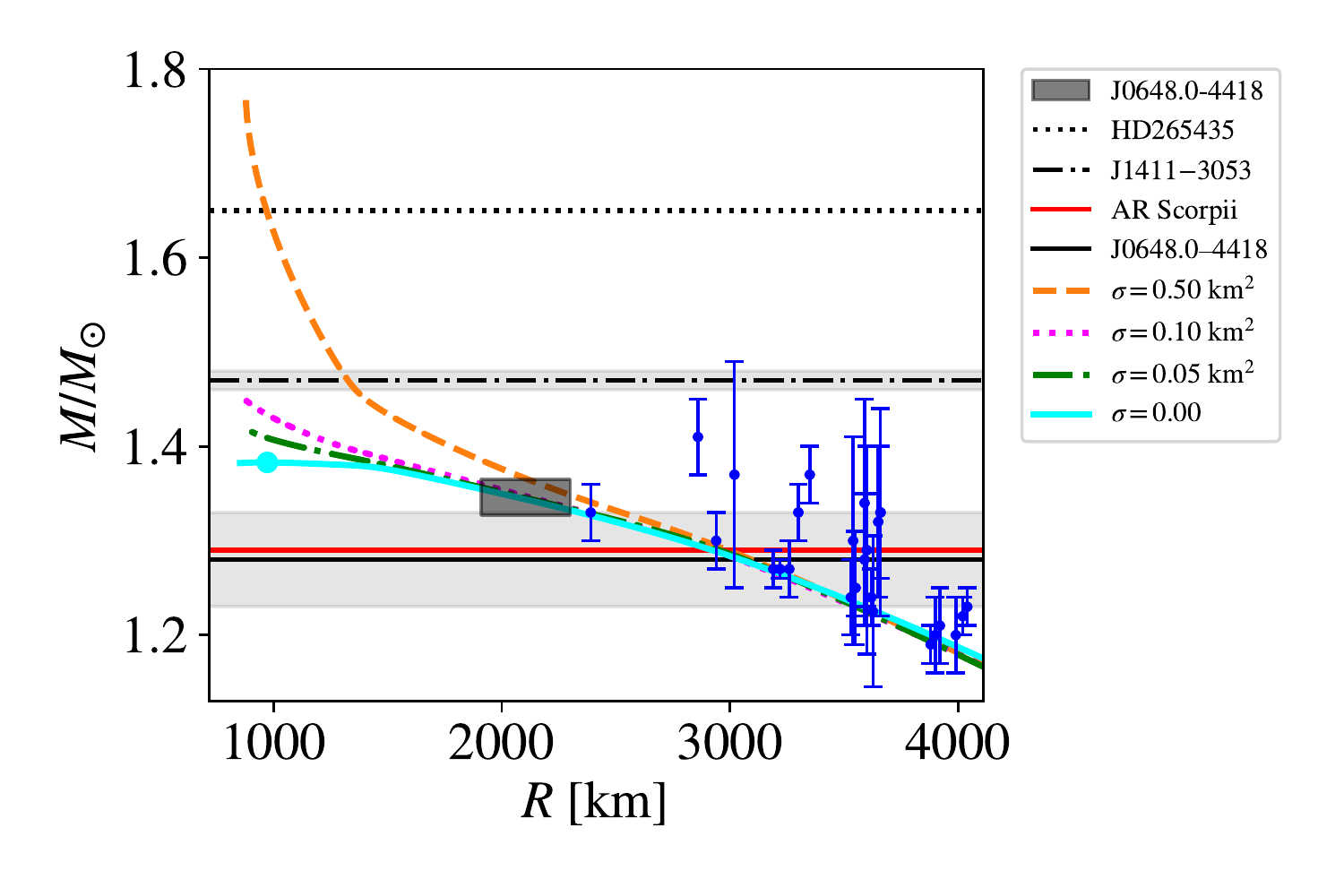}
\caption{Same as Fig.~\ref{fig:he}, but considering the $^{12}$C threshold for equation of state.\label{fig:c}}
\end{figure}

In figure \ref{fig:c}, we have the mass-radius relation for white dwarfs. The threshold for carbon-12 was considered. In this case, the maximum pressure and central
density are: $p_c = 6.99\times 10^{28}\ {\rm dyn\ cm^{-2}}$ and
$\rho_c = 4.20\times 10^{10}\
{\rm g\ cm^{-3}}$. Now, the
influence of the theory on the maximum mass is slightly smaller (the coupling term has dependence on the energy-mass
density), but still strong. It is possible to see, for the highest value of the coupling
constant, $\sigma=0.5\ {\rm km^2}$, that the
maximum mass reaches near $1.8M_{\odot}$. We can also observe that the white dwarfs are more
compact in comparison with the previous case. One can see that the
curves cross the shaded region of RX J0648.0-4418 more to the left
side. This is due to the change in the EoS. For the carbon threshold, the maximum mass limit may be increased if one uses even higher values of $\sigma$.

\begin{figure}[h]
\includegraphics[scale=0.552]{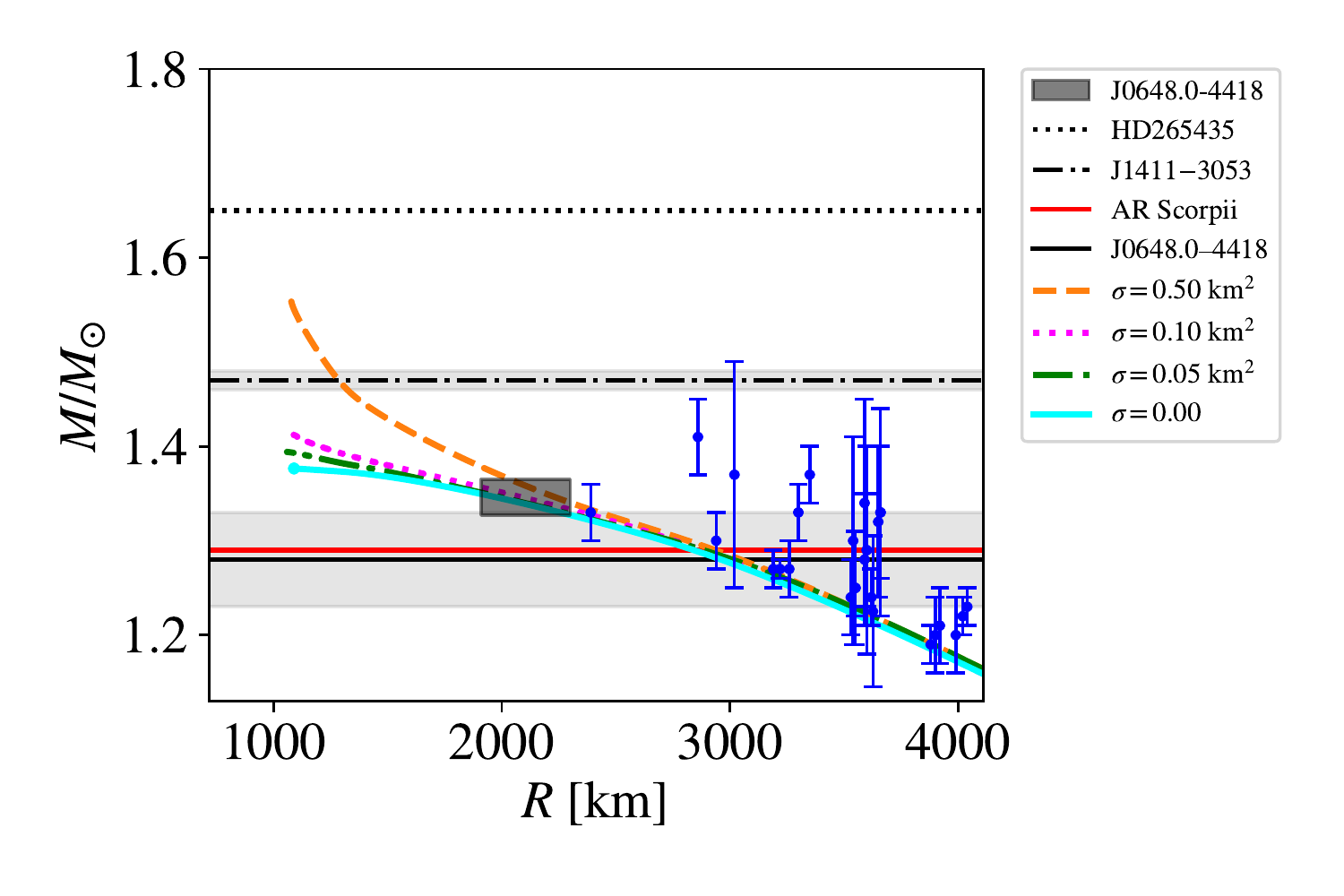}
\caption{Same as Fig.~\ref{fig:he}, but considering the $^{16}$O threshold for equation of state.\label{fig:o}}
\end{figure}

In figure~\ref{fig:o}, we present the mass-radius relationship for white dwarfs considering the
threshold for $^{16}$O, the maximum central pressure is $2.73\times 10^{28}\ {\rm dyn\ cm^{-2}}$, corresponding to a maximum central density of $2.07\times 10^{10}\
{\rm g\ cm^{-3}}$. This
third case follows the same line as the previous one. For the highest coupling constant value, the
maximum mass reached is near $1.6\ M_{\odot}$, which means that smaller density thresholds requires higher values of $\sigma$ to enhance maximum masses up to $2.2-2.4~M_\odot$. The minimum radii
reached are a little more than 1000 km.

\begin{figure}[h]
\includegraphics[scale=0.552]{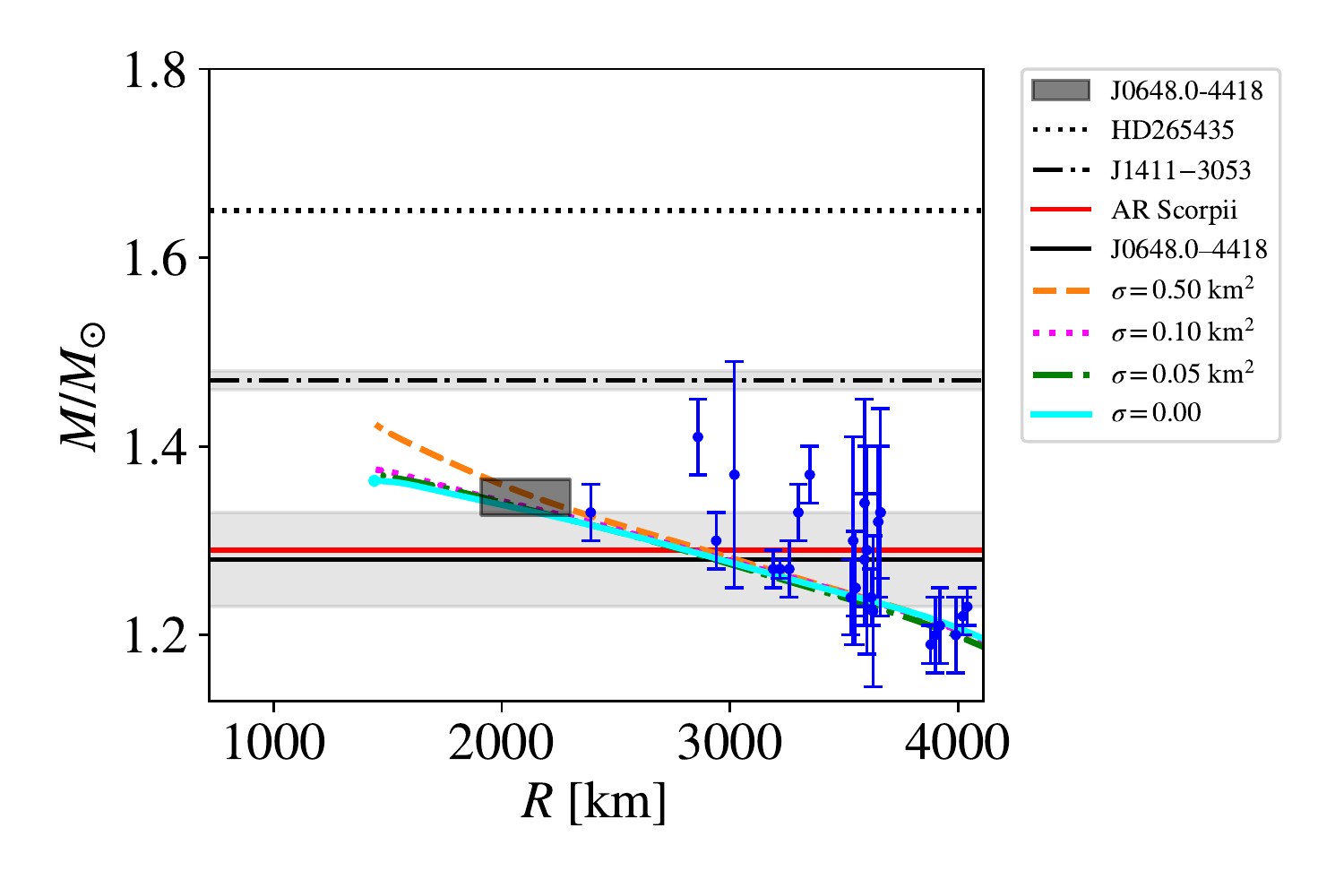}
\caption{Same as Fig.~\ref{fig:he}, but considering the $^{20}$Ne threshold for equation of state.\label{fig:ne}}
\end{figure}

Finally, we present the mass-radius relation for white dwarfs, considering the threshold for $^{20}$Ne in figure \ref{fig:ne}. In
this last case, the maximum pressure and
central density are: $p_c = 6.21\times 10^{27}\ {\rm dyn\ cm^{-2}}$ and
$\rho_c = 6.89\times 10^{9}\ {\rm g\ cm^{-3}}$. It has a maximum mass of
just a few percents above the
Chandrasekhar limit for the highest value of the parameter of the theory
considered. As we can see, for this composition, one cannot have
white dwarfs that would have a mass around $1.47\ M_{\odot}$. The
nuclear instabilities limit a lot the enhancement of the mass for WDs.
Considering elements heavier than oxygen, the nuclear instabilities
will largely affect the maximum mass and limit it before gravitational effects.

\section{Discussion and conclusions}\label{conclu}

In this work, we have studied for the first time the mass-radius
relationship of massive white dwarfs in the context of $\frl$
gravity. We have considered the specific case $\frl = \fr + \fl + \sigma\fr\fl$, where the term
$\sigma\fr\fl$ represents a non-minimal coupling between the matter and gravitational field. We solved the
hydrostatic equilibrium equation for the Hamada-Salpeter EoS and used electron capture instability as
threshold for the central mass density and consequently for the central pressure. We have
found that the effects of the theory are more important for stars with radius less than 2500 km,
i.e., for more massive stars. This is because the coupling term is dependent on the
energy-mass density. Hence, the effects of the theory are noticeable for larger central
densities. We see that curves start to deviate from the General
  Relativity case. We found here that the coupling parameter is at least one order of magnitude smaller
  than it is for neutron stars. This is expected since WDs have a central density smaller than the central
  densities of neutron stars,
  and $\sigma$ has a dependence on the energy-matter density. This behavior is the same as the
  non-minimal model $f(\fr,T)$ of gravity, e.g., see figures in Ref. \cite{moraes/2018a}. As one can see,
the parameters in the theory have different values depending on the astrophysical system.

Furthermore, it is remarkable that the effects of the theory are negligible for stars with $M < 1.3M_{\odot};
R < 3000$ km, which means that the theory recovers General Relativity and Newtonian results
for small densities independent of the value of the coupling constant, i.e.,
the curves are indistinguishable at low densities. This means that
  the theory can accommodate the observational data of WDs with $< 3000$
  km, without any problem. This behavior is different from other
  theories of gravity, which in general do not recover this limit on
  this region. e.g., see figures of Ref. \cite{banerjee/2017}.
 $\frl$ gravity has a similarity with the $f(\fr)$ in the Palatini
  formalism \cite{sarmah/2022},
  where the effects of the alternative theory are negligible in this
  regime. As one goes to higher densities, the curves deviates from
  the GR limit, showing the contribution of the modified gravity on
  the mass-radius relation. This is
  completely in agreement with the data, which shows that for less
massive WDs ($M<1.3M_\odot$), the Newtonian theory can
  describe very well the data. With new data showing more massive WDs
  \cite{nalezyty/2004, tremblay/2011, pshirkov/2020, hollands/2020},
  the relativistic effects, i.e., gravity corrections, are
  necessary to explain the mass-radius.

For stars near the threshold due to the electron-ion interactions, the effects of the theory lead to
stars well above the Chandrasekhar mass limit. For a star made of oxygen, the increment could be at least
$0.18\ M_{\odot}$, for Ne it is $0.06\ M_{\odot}$. The increment
would reduce a lot or be negligible for white dwarfs made of heavier elements, such as $^{40}$Ca or
$^{56}$Fe due to nuclear instabilities. For lighter elements the increment in the mass can be very significant, reaching a limit
more than $2.0\ M_{\odot}$, thus explaining the superluminous supernovae type SNIas. The enhancement in the mass
can be comparable to the ones coming from magnetic field effects \cite{chamel/2013, chamel/2015, otoniel/2017,
  Otoniel2019Jul}, without the anisotropic instabilities of the huge
magnetic field. However, it is worth considering
other nuclear instabilities besides the
electron capture.

\begin{figure}[h]
\includegraphics[scale=0.552]{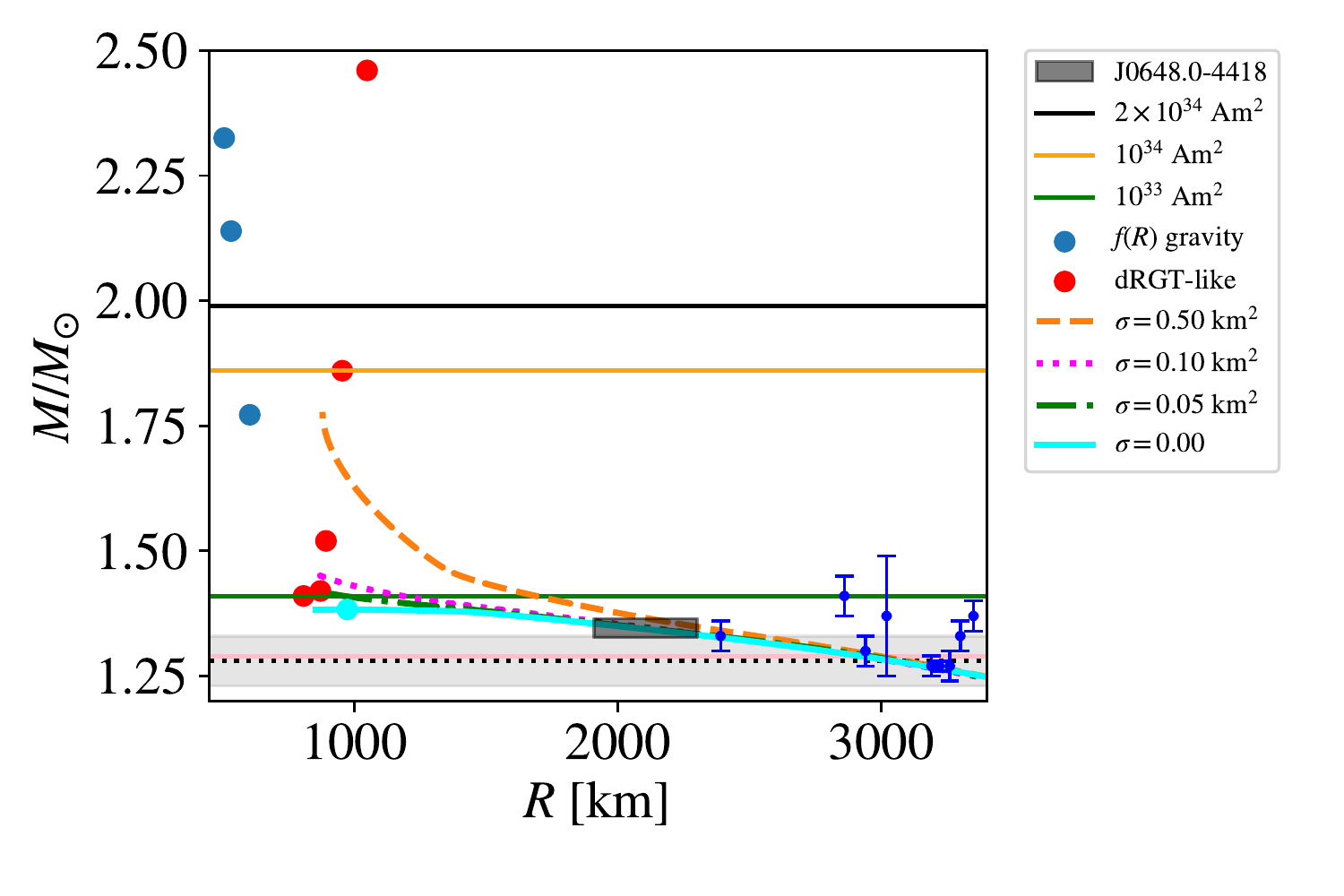}
\caption{Mass radius relationship for white dwarfs with different $\sigma$ parameters and
  considering a star made of $^{12}$C. Four values of $\sigma$ were considered. For $\sigma=0.0$,
  the theory represents General Relativity. The black-shaded region indicates the observed mass-radius
  of ZTFJ1901+1458, this region, i.e., the mass-radius are taken from Ref.~\cite{caiazzo/2021}. The
    black dotted-line and shaded region are $1.28\pm0.05\ M_{\odot}$
    corresponding to RX J0648.0-4418 \cite{mereghetti/2011}. The pink
    line represents the upper limit of the compact object in AR
    Scorpii \cite{marsh/2016}. The blue circles with error bars represent the
  observational data of a sample of massive WDs taken from
  Refs.~\cite{vennes/1997, nalezyty/2004}. In this
  figure we have included the maximum mass of strongly magnetized
  $^{12}$C WDs from Ref. \cite{chatterjee/2017a}. They are represented
by the green, red, and black solid lines. These masses correspond
to the following magnetic moments: $10^{33}, 10^{34}$ and
$2\times10^{34}\ \rm{A m^2}$. We also included two samples of
mass-radius: from Ref. \cite{das/2015}, which considers the $f(\fr)$ gravity and from
Ref. \cite{eslampanah/2019}, which considers the Vegh's massive
gravity. They are represented by the blue and red dots,
respectively.\label{figc:comp}}
\end{figure}

In figure \ref{figc:comp} we compare our results of
  Fig. \ref{fig:c} with the maximum mass of strongly magnetized
  $^{12}$C WDs from Ref. \cite{chatterjee/2017a}. They are represented
by the green, red, and black solid lines where the masses correspond
to the magnetic moments $10^{33}, 10^{34}$ and
$2\times10^{34}\ \rm{A m^2}$ respectively.
The magnetic moment of $10^{33}\ \rm{A m^2}$ corresponds
to a magnetized $^{12}$C WD of $1.41\ M_{\odot}$,
the $10^{34}\ \rm{A m^2}$ corresponds
to $1.86\ M_{\odot}$ and $2\times10^{34}\ \rm{A m^2}$ corresponds
to $1.99\ M_{\odot}$. For the purpose of comparison, when considering $\frl$ gravity effects with $\sigma=0.5\ {\rm km^2}$, we could come close to $1.8\ M_{\odot}$ for
carbon-12 WDs. We also compare with two samples of
mass-radius: one from Ref. \cite{das/2015}, which considers the
$f(\fr)$ gravity and another one from
Ref. \cite{eslampanah/2019}, which considers the Vegh's massive
gravity. They are represented by the blue and red dots,
respectively. As one can see for this case, the modified theories of
$f(\fr)$ and massive gravity can reach more massive stars, however, if
one considers nuclear instabilities the scenario can be thoroughly changed. As we mentioned, the nuclear instabilities limit the maximum
mass significantly, so the WD EoS should always consider the threshold.
Some results for massive gravity are similar to ours.

\begin{figure}[h]
\includegraphics[scale=0.552]{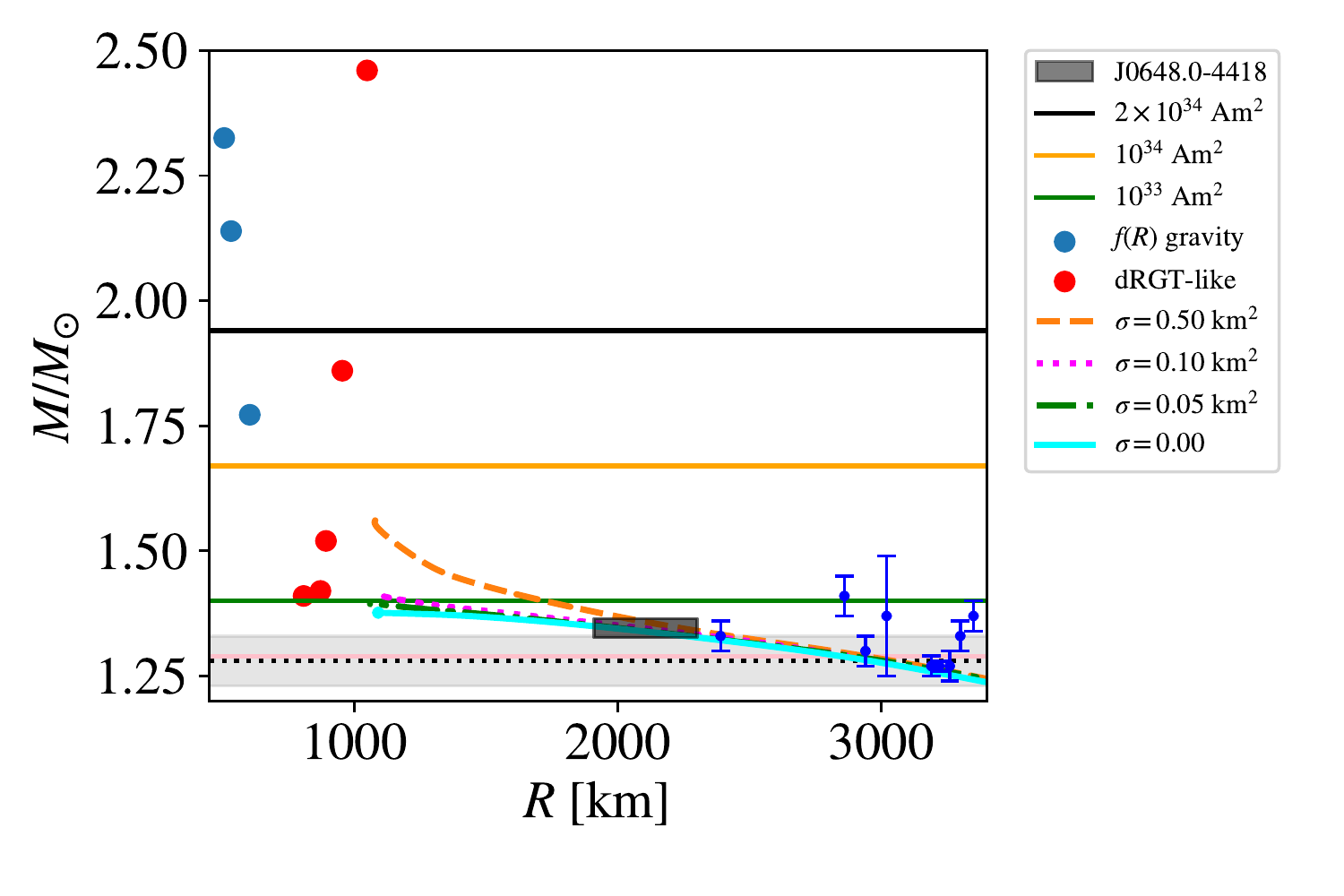}
\caption{Same as Fig.~\ref{figc:comp}, but considering the $^{16}$O
  threshold for equation of state.}
  \label{figo:comp}
\end{figure}

  In figure \ref{figo:comp} we repeat the methodology for WDs made of
oxygen-16, in comparison with the Fig. \ref{fig:o}. As one can see the
maximum mass that considers magnetic field decreases. As we have also
used electron-capture as threshold, our results also decreased, i.e.,
they are consistent for these $^{16}$O WDs and present a very similar
behavior for the maximum mass. As one can see, this modified gravity,
$\frl$ model, can be as good as highly magnetized models.

In conclusion, we show that the $\frl$ modified theory of gravity can explain super-Chandrasekhar white dwarfs,
i.e., a non-minimal coupling between the particle fields and the gravitational field could enhance the maximum
mass for white dwarfs above the Chandrasekhar limit. More aspects of the theory must be
addressed, such as the bounds for the coupling parameter from statistical analyses, to establish a new maximum mass limit. The electron capture limits the maximum central density for different WD compositions
  \cite{chamel/2015, chamel/2016}. One could
  enhance the $\sigma$ parameter, since it reduces the central density for the same star. However,
  there are other instabilities that need to be considered such as
those due to pycnonuclear reactions.
  The theory allows the mass of WDs
  to be higher without surpassing the threshold for nuclear
  instabilities. Another valid consideration would be to have the theory coupled to the electromagnetic field and have magnetic WDs within the theory.

  Presently, we can say
    that electron capture instabilities and estimated masses of
    superluminous SNIas constrain the coupling parameter within the
    interval $0 < \sigma <1\ {\rm km^2}$ for white dwarfs systems.

\begin{acknowledgements}
R.V.L. is supported by U.S. Department of Energy (DOE) under
grant DE--FG02--08ER41533 and to the LANL Collaborative Research Program by Texas A\&M System National
Laboratory Office and Los Alamos National Laboratory and by Uniandes university. G.A.C. is supported by Coordena\c c\~ ao de Aperfei\c coamento de Pessoal de N\'ivel Superior (CAPES) grant PN\-PD/88887.368365/2019-00.
\end{acknowledgements}

\bibliographystyle{spphys}
\bibliography{lib}
\end{document}